\documentclass[10pt,pra,twocolumn,floatfix]{revtex4}
\usepackage{amsfonts}
\usepackage{amssymb}
\usepackage{amsmath}
\usepackage[dvips]{graphicx}
\usepackage{color}

\begin{document}

\title{Experimental test of error-disturbance uncertainty relation with
continuous variables}
\author{Yang Liu$^{1}$}
\author{Haijun Kang$^{1}$}
\author{Dongmei Han$^{1}$}
\author{Xiaolong~Su$^{1,2}$}
\email{suxl@sxu.edu.cn}
\author{Kunchi Peng$^{1,2}$}
\affiliation{$^{1}$State Key Laboratory of Quantum Optics and Quantum Optics Devices, \\
Institute of Opto-Electronics, Shanxi University, Taiyuan, 030006, People's
Republic of China \\
$^{2}$Collaborative Innovation Center of Extreme Optics, Shanxi University,\\
Taiyuan, Shanxi 030006, People's Republic of China\\
}

\begin{abstract}
Uncertainty relation is one of the fundamental principle in quantum
mechanics and plays an important role in quantum information science. We
experimentally test the error-disturbance uncertainty relation (EDR) with
continuous variables for Gaussian states. Two conjugate continuous-variable
observables, amplitude and phase quadratures of an optical mode, are
measured simultaneously by using a heterodyne measurement system. The EDR
with continuous variables for a coherent state, a squeezed state and a
thermal state are verified experimentally. Our experimental results
demonstrate that Heisenberg's EDR with continuous variables is violated, yet
Ozawa's and Branciard's EDR with continuous variables are validated.
\end{abstract}

\maketitle

\section{Introduction}

As one of the cornerstones of quantum mechanics, uncertainty relation
describes the measurement limitation on two incompatible observables.
Uncertainty relation has a huge impact on quantum information technology,
such as entanglement verification \cite{Buscemi}, quantum key distribution 
\cite{Furrer}, quantum dense coding \cite{Bennett} and security of quantum
cryptography \cite{Gisin}. Heisenberg's original uncertainty relation is
related to measurement effect, which states that we cannot acquire perfect
knowledge of a state without disturbing it \cite{Heisenberg}. 

\begin{figure}[tbp]
\begin{center}
\includegraphics[width=80mm]{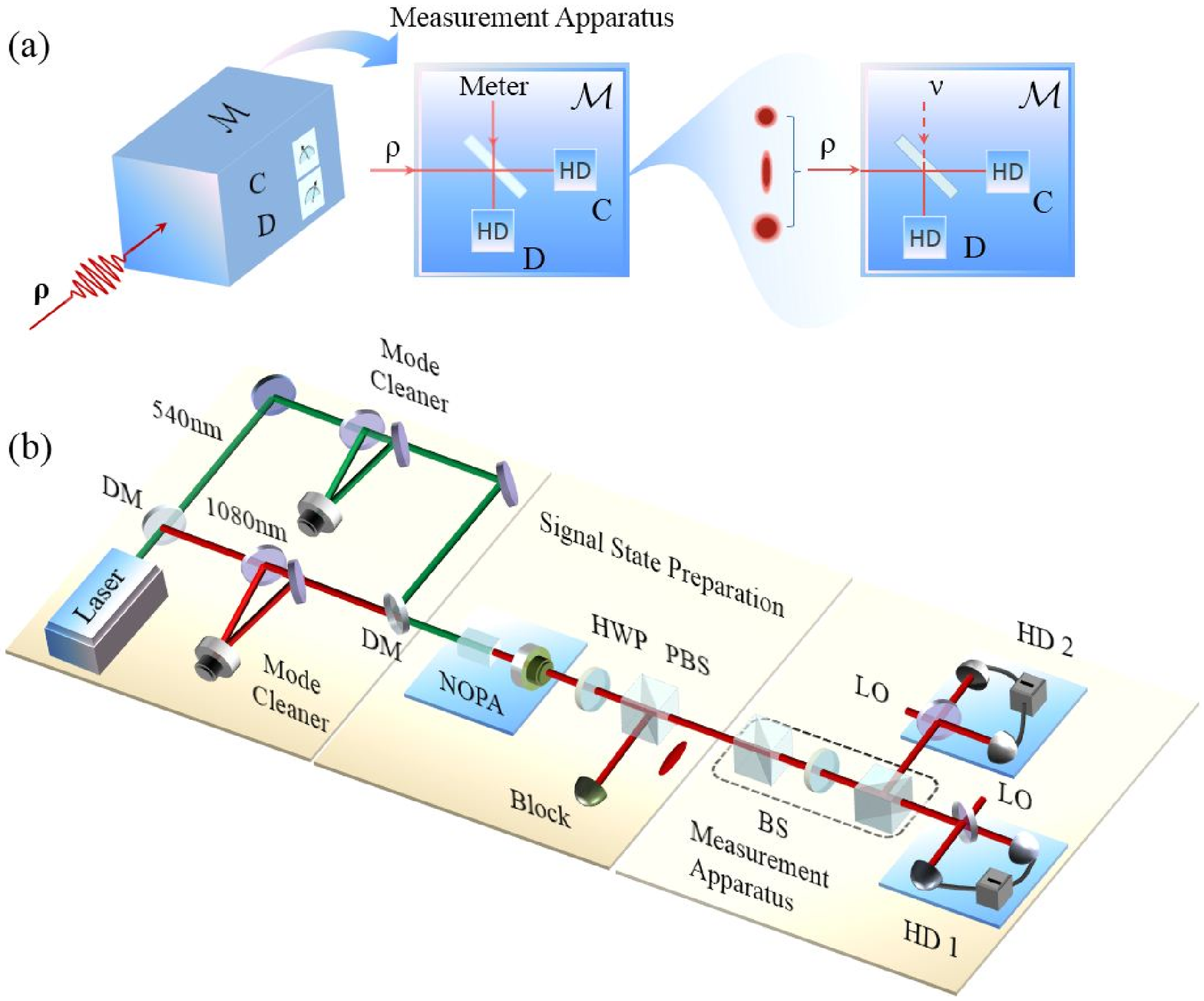}
\end{center}
\caption{(a) The principle of the test of EDR with continuous variables. A
joint measurement apparatus implements the approximation of incompatible
observables $A$ and $B$ with the compatible observables $C$ and $D $ by
coupling the signal mode and meter mode via a beam-splitter. Coherent state,
squeezed state and thermal state serve as signal modes, respectively, and a
vacuum state serves as meter mode. (b) Schematic of the experimental setup.
Signal state is prepared by a NOPA. The measurement apparatus is composed by
a BS, which is a combination of PBS-HWP-PBS, and two HDs. Two output modes
of the BS are detected by HD1 and HD2, respectively. NOPA: nondegenerate
optical parametric amplifier, BS: beam-splitter, HWP: half-waveplate, PBS:
polarization beam-splitter, HD: homodyne detector, LO: local oscillator.}
\end{figure}
There are two kinds of uncertainty relations, which are the preparation
uncertainty relation and the measurement uncertainty relation, depending on
whether you are talking about average measurement or one-shot measurement in
the understanding of Heisenberg's spirit. The preparation uncertainty which
studied the minimal dispersion of two quantum observables before measurement 
\cite{Kennard,Weyl,Rob29}. The Robertson uncertainty relation \cite{Rob29},
reads as $\sigma (x)\sigma (p)\geq \hbar /2$, is a typical example in this
sense, where $\sigma (x)$ and $\sigma (p)$ are the standard deviations of
position and momentum. For such uncertainty relation, the measurements of $x$
and $p$ are performed on an ensemble of identically prepared quantum
systems. The measurement uncertainty relation thinks the Heisenberg's
uncertainty principle should be based on the observer's effect, which means
that measurements of certain system cannot be made without affecting the
system. This kind of uncertainty relation which studies to what extent the
accuracy of a position measurement is related to the disturbance of the
particle's momentum, so is also called the error-disturbance relation (EDR) 
\cite{Ozawa03,Hall04}. 
\begin{figure*}[tbp]
\begin{center}
\includegraphics[width=150mm]{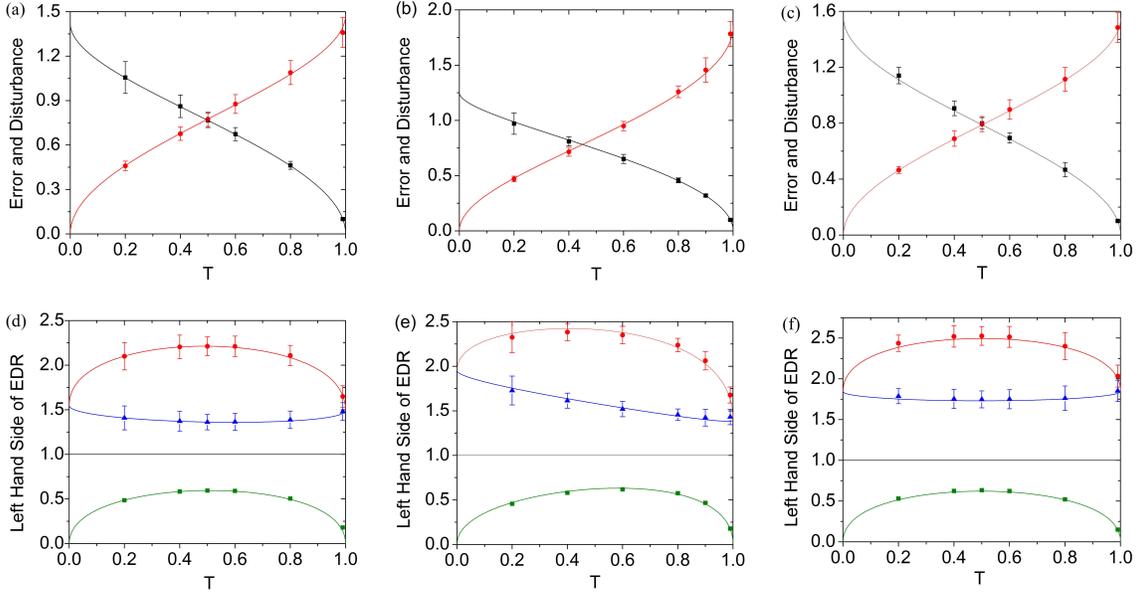}
\end{center}
\caption{Experimental results. (a), (b) and (c): The dependence of error
(black curve) and disturbance (red curve) on the transmission efficiency of
BS (T) for a coherent state, a squeezed state and a thermal state,
respectively. (d), (e) and (f): The left hand sides of the EDRs with
continuous variables for a coherent state, a squeezed state and a thermal
state, respectively. Green curve: the Heisenberg's EDR. Red curve: the
Ozawa's EDR. Blue curve: the Branciard's EDR. Black line: the right hand
side of the EDRs. All experimental data agree well with the theoretical
predictions. The error bars are obtained by RMS of repeated measurement for
ten times.}
\end{figure*}

The Heisenberg's EDR is generally expressed as 
\begin{equation}
\varepsilon (A)\eta (B)\geq C_{AB},
\end{equation}%
where $C_{AB}=\left\vert \left\langle [A,B]\right\rangle \right\vert /2$, $%
[A,B]=AB-BA$, $\varepsilon (A)=\langle (C-A)^{2}\rangle ^{1/2}$ and$\ \eta
(B)=\langle (D-B)^{2}\rangle ^{1/2}$ represent the root-mean-squared (RMS)
difference between the initial value of $A$ and $B$ and the outcome value of
a measurement of $C$ and $D$, respectively. However, it has been shown that
Heisenberg's EDR may be violated in some cases \cite{Balllentine}%
. After that heated debates on EDR have taken place and new formulated of
EDRs have been put forward \cite%
{Ozawa03,Ozawa04,Branciard,Hall04,Werner1,Werner2,PhysRevA022106,PhysRevA032,PhysRevLett050401,lu,Barchielli2017,Barchielli2018}%
. Ozawa proposed the EDR as%
\begin{equation}
\varepsilon (A)\eta (B)+\varepsilon (A)\sigma (B)+\sigma (A)\eta
(B)\geqslant C_{AB}.
\end{equation}%
After that Branciard improved the Ozawa's EDR as \cite{Branciard}%
\begin{eqnarray}
&&\big[\varepsilon ^{2}(A)\sigma ^{2}(B)+\sigma ^{2}(A)\eta ^{2}(B)  \notag
\\
&&+2\varepsilon (A)\eta (B)\sqrt{\sigma ^{2}(A)\sigma ^{2}(B)-C_{AB}^{2}}%
\big]^{1/2}\geqslant C_{AB},
\end{eqnarray}%
which is tighter than Ozawa's EDR. The experimental tests of the uncertainty
relations have been demonstrated in photonic \cite%
{EXPphotons1,EXPphotons2,EXPphotons3,EXPphotons4,EXPphotons5,EXPphotons6},
spin-$^{1}$/$_{2}$ \cite%
{EXPpolarizedneutrons1,EXPpolarizedneutrons2,EXPphotons7,EXPphotons8},
nuclear spin \cite{EXPW1}, and ion trap \cite{EXPW2,EXPW3} systems. All of
these experiments are in discrete-variable systems. Until recently, the test of the error-tradeoff 
uncertainty relation with continuous variables is experimentally demonstrated by using an Einstein-Podolsky-Rosen (EPR)
entangled state \cite{Yang}.

\begin{figure*}[tbp]
\begin{center}
\includegraphics[width=150mm]{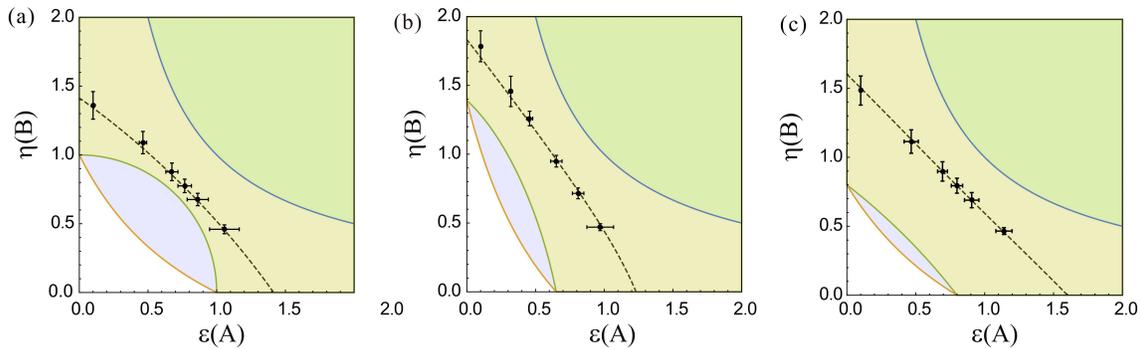}
\end{center}
\caption{Comparison of the lower bounds of EDRs for three Gaussian states.
(a), Coherent state as signal mode. (b), Squeezed state as signal mode. (c),
Thermal state as signal mode. Blue curve: the Heisenberg bound. Orange
curve: the Ozawa bound. Green curve: the Branciard bound. Black circles:
experimental data. Black dotted curve: the theoretical prediction for
experimental parameters. }
\end{figure*}
In this paper, we report the experimental test of EDR with continuous
variables by using a heterodyne measurement system. In our experiment, we
test the EDR for three different Gaussian states, which are coherent state,
squeezed state and thermal state, respectively. A vacuum mode is used as
meter mode in the measurement system. 
Our experimental results demonstrate that Heisenberg's EDR with continuous
variables is violated, yet Ozawa's and Branciard's EDR with continuous
variables are validated.

\section{The principle and experimental setup}

The amplitude and phase quadratures of an optical mode are incompatible
continuous-variable observables and cannot be measured simultaneously. A
heterodyne measurement system, which is a joint measurement apparatus, can
be used to measure the approximation of $A$ and $B$ with the compatible
observables $C$ and $D$ as shown in Fig. 1(a). The signal mode $\rho $ with
incompatible observables $A=\hat{x}_{\rho }$ and $B=\hat{p}_{\rho }$ is
coupled with a meter mode via a beam-splitter (BS), where $\hat{x}=\hat{a}+%
\hat{a}^{\dag }$ and $\hat{p}=(\hat{a}-\hat{a}^{\dag })/i$ denote the
amplitude and phase quadrature of an optical mode, respectively. The signal
mode are prepared as coherent state, squeezed state and thermal state,
respectively, and a vacuum state $\nu $ is used as the meter mode in our
experiment. The amplitude quadrature $C=\hat{x}_{c}=\sqrt{T}\hat{x}_{\rho }-%
\sqrt{R}\hat{x}_{\nu }$ and phase quadrature $D=\hat{p}_{d}=\sqrt{R}\hat{p}%
_{\rho }+\sqrt{T}\hat{p}_{\nu }$ of two output modes $c$ and $d$ of BS are
measured by two homodyne detectors simultaneously, which are used to
approximate\ $A$ and $B$, respectively, where $T$ is the transmission
efficiency of the BS, and $R=1-T$. The root-mean-square error and
disturbance are expressed as

\begin{eqnarray}
&&\varepsilon ^{2}(A)=\langle (C-A)^{2}\rangle  \notag \\
&=&(\sqrt{T}-1)^{2}\sigma ^{2}(\hat{x}_{\rho })+R\sigma ^{2}(\hat{x}_{\nu })
\notag \\
&=&\langle \lbrack (1-\sqrt{T})\hat{x}_{c}-\sqrt{R}\hat{x}_{d}]^{2}\rangle ,
\end{eqnarray}%
\begin{eqnarray}
&&\eta ^{2}(B)=\langle (D-B)^{2}\rangle  \notag \\
&=&(\sqrt{R}-1)^{2}\sigma ^{2}(\hat{p}_{\rho })+T\sigma ^{2}(\hat{p}_{\nu })
\notag \\
&=&\langle \lbrack (1-\sqrt{R})\hat{p}_{c}-\sqrt{T}\hat{p}_{d}]^{2}\rangle .
\end{eqnarray}

The experimental setup for test of EDR is illustrated in Fig. 1(b). A laser
generates both 1080 nm and 540 nm optical fields simultaneously. The 1080 nm
optical field is used as the injected signal of a nondegenerate optical
parametric amplifier (NOPA) and the local oscillator fields of homodyne
detectors. The 540 nm optical field serves as the pump field of the NOPA. A
half-waveplate (HWP) and a polarization beam-splitter (PBS), which are
placed after the NOPA, are used to obtain different signal modes. The
measurement apparatus is composed by a BS and two homodyne detectors. The AC
output signals from HD1 and HD2 are mixed with a local reference signal of 3
MHz, and then filtered by low-pass filter with a bandwidth of 30 kHz and
amplified 1000 times (Low noise preamplifier, SRS, SR560), respectively. And
then the two signals from the outputs of the preamplifiers are recorded by a
digital storage oscilloscope simultaneously. A sample size of 5$\times 10^{5%
\text{ }}$data points is used for all quadrature measurements. The
interference efficiency between signal and local oscillator fields is 99\%
and the quantum efficiency of photodiodes are 99.6\%.


\section{Results}

A coherent state is prepared when the pump field of NOPA is blocked and only
the injected field passes through the NOPA. The variances of amplitude and
phase quadratures of the coherent state and the vacuum state (meter mode)
are $\sigma ^{2}(\hat{x}_{\rho })=\sigma ^{2}(\hat{p}_{\rho })=1$, and $%
\sigma ^{2}(\hat{x}_{\nu })=\sigma ^{2}(\hat{p}_{\nu })=1$, respectively.
When the NOPA is operated at the parametric deamplification situation and
the half-waveplate after the NOPA is set to $22.5^{\circ }$, an x-squeezed
and a p-squeezed states are prepared. The x-squeezed state is used as the
signal mode in the test of EDR for squeezed state. The variances of the
amplitude and phase quadratures of the x-squeezed state are $\sigma ^{2}(%
\hat{x}_{\rho })=e^{-2r},\sigma ^{2}(\hat{p}_{\rho })=e^{2r}$, respectively,
where $r$\ is the squeezing parameter \cite{EPRSU}. In the experiment, the
squeezed state with -2.9 dB squeezing and 3.9 dB antisqueezing is generated
by NOPA. When the half-waveplate after the NOPA is set to $0^{\circ }$, the
Einstein-Podolsky-Rosen entangled state is generated. Each mode of the
entangled state is a thermal state, and one of them is used for the test of
EDR for thermal state. The variances of the amplitude and phase quadratures
of the thermal state are $\sigma ^{2}(\hat{x}_{\rho })=\sigma ^{2}(\hat{p}%
_{\rho })=(e^{-2r}+e^{2r})/2$.

The dependence of error of the amplitude quadrature $\varepsilon (A)$ and
disturbance of the phase quadrature $\eta (B)$\ on the transmission
efficiency of BS for three different Gaussian signal modes are shown in Fig.
2(a), 2(b) and 2(c), respectively. The error $\varepsilon (A)$ decreases
with the increasing of the transmission efficiency of the BS, while the
disturbance $\eta (B)$\ increases with the increasing of the transmission
efficiency for all of the three Gaussian states. When the error reaches
minimum value, the maximum disturbance is caused. The reduction of the
disturbance can be realized by introducing error on the other observable.
When a x-squeezed state serves as signal mode, the maximum error is less
than the case that coherent state serves as signal field with the cost of
the greater maximum disturbance for the anti-squeezing of the phase
quadrature [Fig. 2 (b)]. When the signal mode is a thermal state, both the
error and disturbance of the state are larger than that of coherent state at
the same transmission efficiency of BS, as shown in Fig. 2(a) and 2(c).

The dependence of the left hand side of Ozawa's (red curve), Branciard's\
(blue curve) and Heisenberg's (green curve) EDRs with continuous variables
on the transmission efficiency of BS for three Gaussian states are shown in
Fig. 2(d), 2(e) and 2(f), respectively. It is clear that the Ozawa's and
Branciard's EDR with continuous variables are valid while the Heisenberg's
EDR with continuous variable is violated. Comparing the blue curve and red
curve, we can see that the Branciard's EDR is tighter than Ozawa's EDR with
continuous variables. When the transmission efficiency is 50\%, the left
hand side of Branciard's EDR with continuous variables reaches its minimum
value in case of coherent state and thermal state. In the case of x-squeezed
state serves as signal mode, the Branciard's inequality is minimized when
the transmission efficiency is about 95\% for the unsymmetrical of the
variances of amplitude and phase quadratures of the squeezed state.

The comparison of the lower bounds of EDRs for three Gaussian states in the
error-disturbance plot are shown in Fig. 3. The results for coherent state,
squeezed state and thermal state are shown in Fig. 3(a), 3(b), and 3(c),
respectively. All the experimental results demonstrate that Heisenberg's EDR
is violated, yet Ozawa's and Branciard's EDR with continuous variables are
valid.

\section{Conclusion}

We experimentally test the Heisenberg's, Ozawa's and Branciard's EDRs with
continuous variables by using a heterodyne measurement system. Three
different Gaussian states, i.e., coherent state, squeezed state and thermal
state are used as signal mode to test the EDRs. All the experimental results
demonstrate that Heisenberg's EDR is violated, yet Ozawa's and Branciard's
EDR are validated. Our work represents an important advance in understanding
fundamentals of physical measurement and sheds light on the developing of
quantum information technology.

\section*{ACKNOWLEDGMENTS}

This research was supported by the NSFC (Grant No. 11834010), the program of
Youth Sanjin Scholar, the National Key R\&D Program of China (Grant No.
2016YFA0301402), and the Fund for Shanxi "1331 Project" Key Subjects
Construction.


\begin{thebibliography}{99}
\bibitem{Buscemi} F. Buscemi, All Entangled Quantum States Are Nonlocal,\
Phys. Rev. Lett. \textbf{108,} 200401 (2012).

\bibitem{Furrer} F. Furrer, T. Franz, M. Berta, A. Leverrier, V. B. Scholz,
M. Tomamichel, and R. F. Werner, Continuous Variable Quantum Key
Distribution: Finite-Key Analysis of Composable Security against Coherent
Attacks,\ Phys. Rev. Lett. \textbf{109}, 100502 (2012).

\bibitem{Bennett} C. H. Bennett and S. J. Wiesner, Communication via one-
and two-particle operators on Einstein-Podolsky-Rosen states,\ Phys. Rev.
Lett. \textbf{69}, 2881 (1992).

\bibitem{Gisin} N. Gisin, G. Ribordy, W. Tittel, and H. Zbinden, Quantum
cryptography,\ Rev. Mod. Phys. \textbf{74}, 145 (2002).

\bibitem{Heisenberg} W. Heisenberg,\textit{\ }\"{U}ber den anschaulichen
Inhalt der quantentheoretischen Kinematik und Mechanik,\ Z. Phys\textit{. }%
\textbf{43},\textbf{\ }172 (1927).

\bibitem{Kennard} E. H. Kennard, Zur Quantenmechanik einfacher
Bewegungstypen,\ Z. Phys. \textbf{44}, 326--352 (1927).

\bibitem{Weyl} H. Weyl, Gruppentheorie und Quantenmechanik (The University
of California, 1928).

\bibitem{Rob29} H. P. Robertson, The uncertainty principle,\textit{\ }Phys.
Rev. \textbf{34, }163 (1929).

\bibitem{Ozawa03} M. Ozawa, Universally valid reformulation of the
Heisenberg uncertainty principle on noise and disturbance in measurements,\
Phys. Rev. A \textbf{67,} 042105 (2003).

\bibitem{Hall04} M. J. W. Hall, Prior information: How to circumvent the
standard joint-measurement uncertainty relation,\ Phys. Rev. A\textit{\ }%
\textbf{69,} 052113 (2004).

\bibitem{Balllentine} L. E. Ballentine, The statistical interpretation of
quantum mechanics,\ Rev. Mod. Phys. \textbf{42}, 358--381 (1970).

\bibitem{Ozawa04} M. Ozawa, Uncertainty relations for joint measurements of
noncommuting observables,\ Phys. Lett. A \textbf{320,} 367 (2004).

\bibitem{Branciard} C. Branciard, Error-tradeoff and error-disturbance
relations for incompatible quantum measurements,\ Proc. Natl. Acad. Sci. 
\textbf{110, }6742 (2013).

\bibitem{Werner1} P. Busch, P. Lahti, and R. F. Werner, Heisenberg
uncertainty for qubit measurements,\ Phys. Rev. A \textbf{89}, 012129 (2014).

\bibitem{Werner2} P. Busch, P. Lahti, and R. F. Werner, Colloquium: Quantum
root-mean-square error and measurement uncertainty relations,\ Rev. Mod.
Phys. \textbf{86}, 1261--1281 (2014).

\bibitem{PhysRevA022106} J. Dressel and F. Nori, Certainty in Heisenberg's
uncertainty principle: Revisiting definitions for estimation errors and
disturbance,\ Phys. Rev. A \textbf{89,} 022106 (2014).

\bibitem{PhysRevA032} K. Baek, T. Farrow, and W. Son, Optimized entropic
uncertainty relation for successive measurement,\ Phys. Rev. A\textit{\ }%
\textbf{89, }032108 (2014).

\bibitem{PhysRevLett050401} F. Buscemi, M. J. W. Hall, M. Ozawa, and M. M.
Wilde, Noise and disturbance in quantum measurements: An
information-theoretic approach,\ Phys. Rev. Lett.\textit{\ }\textbf{112,}
050401 (2014).

\bibitem{lu} X. M. Lu, S. Yu, K. Fujikawa, and C. H. Oh, Improved
error-tradeoff and error-disturbance relations in terms of measurement error
components,\ Phys. Rev. A \textbf{90,} 042113 (2014).

\bibitem{Barchielli2017} A. Barchielli, M. Gregoratti, and A. Toigo,
Measurement uncertainty relations for position and momentum: Relative
entropy formulation,\ Entropy \textbf{19,} 301 (2017).

\bibitem{Barchielli2018} A. Barchielli, M. Gregoratti, and A. Toigo,
Measurement Uncertainty Relations for Discrete Observables: Relative Entropy
Formulation,\ Communications in Mathematical Physics, \textbf{357},
1253--1304 (2018).

\bibitem{EXPphotons1} M. Ringbauer, D. N. Biggerstaff, M. A. Broome, A.
Fedrizzi, C. Branciard, and A. G. White, Experimental joint quantum
measurements with minimum uncertainty,\ Phys. Rev. Lett. \textbf{112},
020401 (2014).

\bibitem{EXPphotons2} F. Kaneda, S.Y. Baek, M. Ozawa, and K. Edamatsu,
Experimental test of error-disturbance uncertainty relations by weak
measurement,\ Phys. Rev. Lett. \textbf{112}, 020402 (2014).

\bibitem{EXPphotons3} L. A. Rozema, A. Darabi, D. H. Mahler, A. Hayat, Y.
Soudagar, and A. M. Steinberg, Violation of Heisenberg's
measurement-disturbance relationship by weak measurements,\ Phys. Rev. Lett. 
\textbf{109}, 100404 (2012).

\bibitem{EXPphotons4} A. P. Lund and H. M. Wiseman, Measuring
measurement--disturbance relationships with weak values,\ New J. Phys. 
\textbf{12}, 093011 (2010).

\bibitem{EXPphotons5} S. Y. Baek, F. Kaneda, M. Ozawa, and K. Edamatsu,
Experimental violation and reformulation of the Heisenberg's
error-disturbance uncertainty relation,\ Sci. Rep. \textbf{3}, 2221 (2013).

\bibitem{EXPphotons6} M. M. Weston, M. J. W. Hall, M. S. Palsson, H. M.
Wiseman, and G. J. Pryde, Experimental test of universal complementarity
relations,\ Phys. Rev. Lett. \textbf{110}, 220402 (2013).

\bibitem{EXPpolarizedneutrons1} J. Erhart, S. Sponar, G. Sulyok, G. Badurek,
M. Ozawa, and Y. Hasegawa, Experimental demonstration of a universally valid
error--disturbance uncertainty relation in spin measurements,\ Nat. Phys. 
\textbf{8}, 185--189 (2012).

\bibitem{EXPpolarizedneutrons2} G. Sulyok, S. Sponar, J. Erhart, G. Badurek,
M. Ozawa, and Y. Hasegawa, Violation of Heisenberg's error-disturbance
uncertainty relation in neutron-spin measurements,\ Phys. Rev. A \textbf{88}%
, 022110 (2013).

\bibitem{EXPphotons7} G. Sulyok, S. Sponar, B. Demirel, F. Buscemi, M. J. W.
Hall, M. Ozawa, and Y. Hasegawa, Experimental Test of Entropic
Noise-Disturbance Uncertainty Relationsfor Spin-1/2 Measurements,\ Phys.
Rev. Lett. \textbf{115}, 030401 (2015)

\bibitem{EXPphotons8} B. Demirel, S. Sponar, G. Sulyok, M. Ozawa, and Y.
Hasegawa, Experimental Test of Residual Error-Disturbance Uncertainty
Relations for Mixed Spin-1/2 States,\ Phys. Rev. Lett. \textbf{117}, 140402
(2016)

\bibitem{EXPW1} W. Ma, Z. Ma, H. Wang, Z. Chen, Y. Liu, F. Kong, Z. Li, X.
Peng, M. Shi, F. Shi, S. Fei, and J. Du,\textit{\ }Experimental test of
Heisenberg's measurement uncertainty relation based on statistical
distances,\ Phys. Rev. Lett. \textbf{116,} 160405 (2016).

\bibitem{EXPW2} F. Zhou, L. Yan, S. Gong, Z. Ma, J. He, T. Xiong, L. Chen,
W. Yang, M. Feng, and V. Vedral,\textit{\ }Verifying Heisenberg's
error-disturbance relation using a single trapped ion,\ Sci. Adv.\textit{\ }%
\textbf{2,} e1600578 (2016).

\bibitem{EXPW3} T. Xiong, L. Yan, Z. Ma, F. Zhou, L. Chen, W. Yang, M. Feng,
and P. Busch, Optimal joint measurements of complementary observables by a
single trapped ion,\ New J. Phys. \textbf{19} 063032 (2017).

\bibitem{Yang} Y. Liu, Z. Ma, H. Kang, D. Han, M. W, Z. Qin, X. Su, and K.
Peng, \textquotedblleft Experimental test of error-tradeoff uncertainty
relation using a continuous-variable entangled state,\textquotedblright\
arXiv: 1905.05632v1.

\bibitem{EPRSU} X. Su, S. Hao, X. Deng, L. Ma, M. Wang, X. Jia, C. Xie, and
K. Peng, \textquotedblleft Gate sequence for continuous variable one-way
quantum computation,\textquotedblright\ Nat. Commun. \textbf{4,} 2828 (2013).
\end{thebibliography}
\end{document}